\documentclass[12pt,preprint]{aastex}

\slugcomment{Accepted by ApJ}

\shorttitle{Gaseous Fuel}
\shortauthors{Putman, M.E.}

\def\gtrapprox{\;\lower 0.5ex\hbox{$\buildrel >\over \sim\ $}}
\def\lessapprox{\;\lower 0.5ex\hbox{$\buildrel < \over \sim\ $}}

\def\Msun  {${\rm M}_\odot$}
\def\deg   {$^\circ$}

\def\kms   {\ km s$^{-1}$}

\begin{document}

\title{Potential Condensed Fuel for the Milky Way}

\author{M. E. Putman\altaffilmark{1}}
\altaffiltext{1}{Department of Astronomy, University of Michigan, 830 Dennison, 500 Church St, Ann Arbor, MI 48109; mputman@umich.edu}

\begin{abstract}
Potential condensed clouds of gas in the Galactic halo are examined in the context of the recent
models of cooling, fragmenting clouds building up the baryonic mass of the Galaxy.
582 high-velocity clouds (HVCs) are defined as the potential infalling, condensed
clouds and the sample's spatial and velocity distribution are presented.   With
the majority of the hydrogen in the clouds ionized ($\sim 85$\%), the
clouds at a distribution of distances within 150 kpc, and their 
individual total masses below $10^{7}$ \Msun, the total mass in potentially condensed clouds is  
$1.1 - 1.4 \times 10^{9}$ \Msun.  If the tighter distance constraint
of $< 60$ kpc is adopted this mass range drops to $4.5 - 6.1 \times 10^{8}$ \Msun.
%JUST ADDED SOME SO THE RANGE INCLUDES THE EXTRAPOLATED CATALOG
%If the flux distribution function for HVCs is extrapolated to lower fluxes, the total mass
%in potentially condensed clouds that will be detected in the Galactic halo does not substantially change.
%The total mass in halo clouds is thus at least an order of magnitude lower than expected by some condensing
%halo cloud models.   
%Additional scenarios for the distribution of clouds are considered and explanations for the discrepancy 
%with models, such as rapid cloud infall, 
The implications on the condensing cloud models, as well as feedback and additional accretion methods, are discussed. 
%It is possible a large fraction of the clouds have not had time to cool and form neutral hydrogen,
%are ionized from the extragalactic ionizing radiation field, and/or have not yet been 
%detected due to their low column densities.   
\end{abstract}

\keywords{Galaxy: halo $-$ intergalactic medium $-$ galaxies: formation $-$ cooling flows $-$ Galaxy: formation}

\section{Introduction}

The range of stellar ages and metallicities in
galaxies like the Milky Way indicate fresh star 
formation fuel must fall in throughout their lives (e.g., Rocha-Pinto et al. 2000; Renda et al. 2005).
The gas accretion process has traditionally been thought to proceed
via shock-heated halo gas from the intergalactic medium cooling and falling in to feed the star
formation process (e.g., White \& Rees 1978; White \& Frenk 1991).
Recently these models have been extended from all of the gas within a certain
radius collapsing monolithically,
to including fragmentation as the hot gas cools, forming pressure
supported warm clouds (e.g., Maller \& Bullock 2004, hereafter MB04; Kaufmann et al. 2005; Sommer-Larsen 2006).
Models which include fragmentation do not have the "over-cooling problem" the monolithic
collapse models have.  In another words, all of the gas does not cool and fall in at once, a large
fraction remains in the halo in a warm/hot phase and excessive feedback is not necessary to explain
the observed baryonic mass of the galaxies.
MB04 predict at any given time during a Milky Way-like galaxy's recent
evolution, several thousand condensed clouds with a total mass on the order of $2 \times 10^{10}$ \Msun~can be found 
within a $\sim$150 kpc radius.  These clouds are pressure confined by the hot gaseous halo
remaining and should currently be found in the
Galactic halo as evidence of this ongoing gas accretion. 

Likely candidates for these condensed infalling clouds are
the high-velocity clouds (HVCs) of neutral hydrogen surrounding our Galaxy.
Oort (1966) was the first to propose this type of origin for HVCs and
MB04 discuss the similarities between the HVCs and the condensed
clouds in their model.   HVCs range in size from ultra-compact ($< 20$ arcmin$^{2}$; e.g., Br\"uns \& 
Westmeier 2004)
to hugely extended (1800 deg$^{2}$; Wakker \& van Woerden 1997) and have typical peak column densities 
of approximately $10^{19}$ cm$^{-2}$ (Putman et al. 2002).  Their velocities generally 
extend from 90 \kms $< |V_{LSR}| <$ 450 \kms, or $-300$ \kms $< V_{GSR} < 300$ \kms.  
Recent progress on the distances to HVCs
allows origins such as the condensing cloud model to be seriously considered and constrained.  
%When considering these origins the large number of
%clouds which are known to originate from the interaction of the Magellanic
%Clouds with the Milky Way should be excluded (e.g. Putman et al. 2003).  
%These clouds should be excluded from any model suggesting
%the clouds are condensing from the hot halo medium.  Secondly, there is a large number of positive, as well as negative velocity
%clouds relative to the Galactic Standard of Reference (GSR).  Though we are only able to measure
%a single velocity component for HVCs, these positive velocity clouds are difficult to 
%put into the infalling condensing cloud models.   Finally, the 
%Recent constraints on the distances
%to many clouds should be considered (approximate distance constraints are in parentheses).  
The direct distance constraints involve looking for absorption 
lines in the spectra of halo stars at known distance and generally provide lower
limits on the order of $> 5$ kpc, but also include upper limits for 3 clouds of $< 10$ kpc (summarized by 
Wakker 2001; Thom et al. 2006).  Indirect distance constraints include deep HI observations of systems 
similar to the Milky Way and
Local Group ($< 160$ kpc; e.g., Pisano et al. 2004; Zwaan 2001), H$\alpha$ observations 
indicating the HVCs
are being ionized by photons escaping from the Milky Way ($< 40$ kpc for some clouds; 
e.g., Putman et al. 2003a), 
and constraints on the properties of the compact HVCs (CHVCs;  $\theta < 1$\deg) when subject to the extragalactic ionizing
background radiation ($< 200$ kpc; e.g., Maloney \& Putman 2003).  In addition, recent
work surveying the environment of M31 does not find the M31 analogs of CHVCs beyond 60 kpc 
(Westmeier, Br\"uns \& Kerp 2005).   All of these distance constraints place
the clouds within the extended Galactic halo and appear to be consistent
with the distances expected for the condensing, infalling clouds in the models.

Given the recent developments both theoretically and observationally regarding  
the gaseous distribution about galaxies, 
the time is right to address the properties of potential condensed 
clouds currently in the Galactic halo.
This paper addresses the observational constraints on the HVC population and how this
can be put together with the models to form a consistent picture of condensed, infalling clouds feeding the Milky Way's star formation.
The selection of high-velocity clouds is presented in the next section, followed by the 
properties of this sample in the context of the condensed cloud model.   
Finally, the results are discussed and summarized.

\section{Data: HVC Selection}

Potential condensed clouds were selected from an updated version of the all-sky HVC catalog of 
Wakker \& van Woerden (1991; hereafter WvW91).
The catalog has been updated by including clouds from the catalog
of HVC components by Morras et al. (2000) for declinations $< -23$\deg\ and
includes clouds with $|V_{\rm LSR}| > 90$ \kms~(see Wakker 2004 for more information on
the catalog properties).
The updated WvW91 catalog of 626 clouds
is not exactly high resolution (0.5\deg~at best), but covers the entire sky
with a detection limit of $2-3 \times 10^{18}$ cm$^{-2}$ ($\Delta$v = 25 \kms).
%looks like both areas are a 0.5deg beam on a 1x1deg grid
The only selection criteria for the HVCs to be potentially condensed halo clouds was the exclusion of clouds 
associated with the Magellanic System (e.g., the Magellanic Stream 
and Leading Arm; Putman et al. 2003b) and the Outer Arm Complex.  The Outer Arm Complex
is a large low latitude structure that is consistent with being a warped
section of the outer Galactic Disk, or a high-z spiral arm (e.g., Wakker \& van Woerden 1997).  
The remaining 582 clouds have a mean velocity relative to the Galactic Standard of Rest of V$_{GSR}$ = -43 \kms. 
The clouds in the condensing cloud models are predicted to have a range of velocities as they form and
move through the halo, but should have a net infall in agreement with this mean V$_{GSR}$ (Bullock pers. comm.).  
The clouds have a total HI flux of 993,557 Jy \kms.   The largest contributors
to this total flux are Complex C (209,590 Jy \kms) and then Complex H (98,040 Jy \kms).
% (see WvW 1991; 1998).  

\section{Results}

\begin{figure}
\includegraphics[angle=90, scale=0.7]{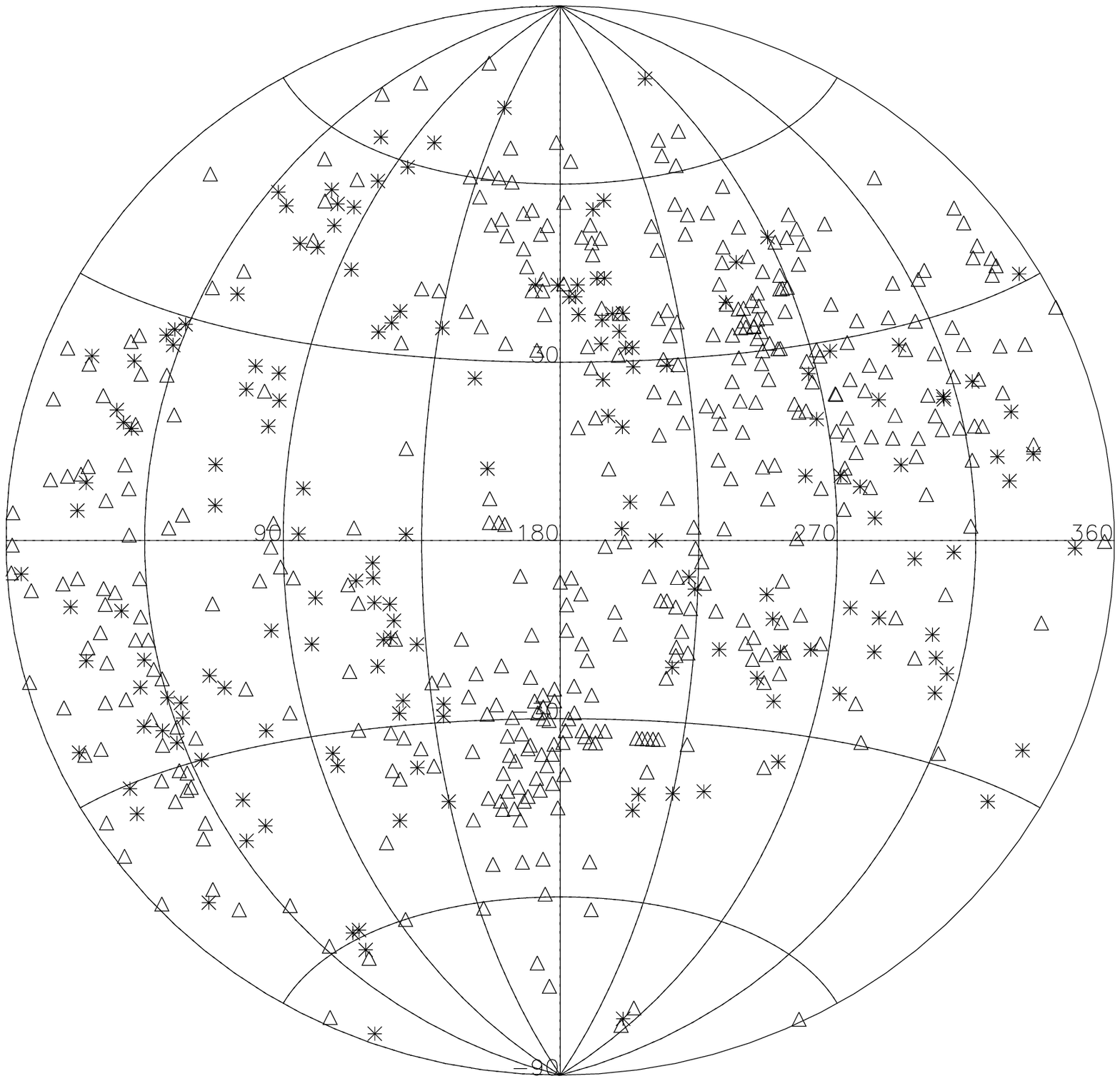}
\caption{Spatial distribution of the 582 potentially condensed high velocity clouds on the sky in Galactic coordinates with
the negative V$_{\rm GSR}$ clouds labeled as triangles and the positive V$_{\rm GSR}$ clouds labeled as stars.\label{spatial}}
\end{figure}

The spatial distribution of the potential infalling HVCs is shown in Figure~\ref{spatial} with
the symbol representative of a positive (star) or negative (triangle) V$_{\rm GSR}$ cloud.
The clouds have V$_{\rm GSR}$'s between approximately $+/-$ 300~\kms.
The clouds are distributed throughout the sky with positive and
negative velocity clouds intermingled.
The largest concentrations of clouds are towards the anti-center in the southern 
Galactic hemisphere and around $l = 260$\deg~  in the northern Galactic hemisphere.
These clouds are thus found looking away from 
the majority of Galactic disk, however there is also a number of clouds found in
both hemispheres from $l = 0-45$\deg.  
The clouds with V$_{\rm GSR}$ more negative than -100 \kms~ are also
concentrated between  $l = 0-45$\deg~ and around $l = 180$\deg.

\begin{figure}
\includegraphics[angle=90, scale=0.9]{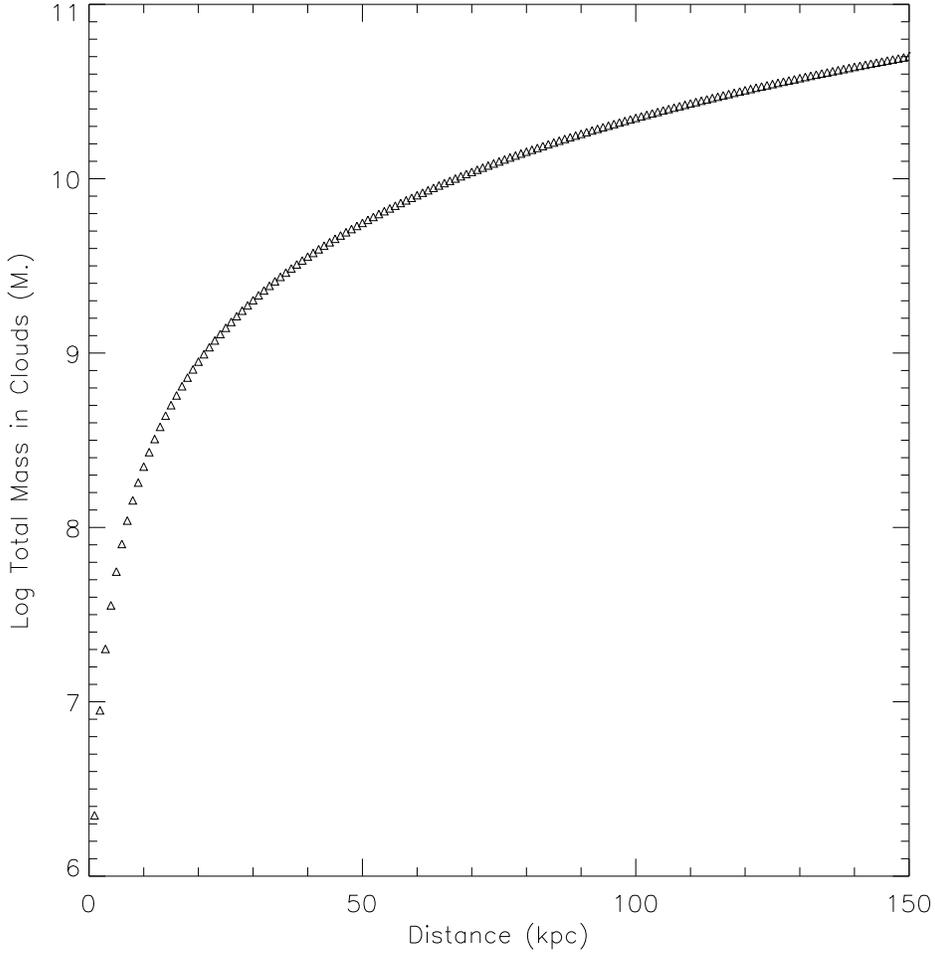}
\caption{The total mass of the entire population of potentially infalling clouds if they are
all given the typical distance on the x axis and are assumed to be 10.5\% observable
neutral hydrogen (see text).  The HI flux used to calculate this total mass is dominated
by some of the large complexes like Complex C, so it is more realistic
to place the clouds at a distribution of distances which keeps the individual cloud masses
below $10^{7}$ \Msun~ as discussed in the text. \label{himass}}
\end{figure}

The total mass of the entire population of potentially infalling clouds as a function of average distance is
shown in Figure~\ref{himass}.  This plot assumes the fraction of the cloud that is detectable
as neutral hydrogen is 10.5\% of the total mass of the cloud; 70\% of the cloud is hydrogen
and 15\% of that hydrogen has cooled to be detectable as HI.   This 15\% is justified
by assuming the extragalactic ionizing radiation field is the dominant factor in
ionizing the clouds (e.g., Maloney \& Putman 2003; see Section 4.3).  
The HI mass was calculated using, M$_{HI} = 2.36 \times 10^{5}$ I (Jy\kms) D$^{2}$(Mpc),
with I = 993,557 Jy\kms~for all of the clouds selected according
to the previous section.  The total mass, M$_{tot}$, is then M$_{HI}/0.105$.
With this constraint, Figure 2 shows that if the high velocity HI flux is at an average distance of 80 kpc the 
total cloud mass is $\sim2 \times 10^{10}$ \Msun, at 40 kpc the total mass would be $\sim6 \times 10^{9}$ \Msun, 
and $\sim9 \times 10^{8}$ \Msun\ at an average distance of 20 kpc.  

There are several clouds that contribute significantly to the total HI flux in halo clouds of 993,557 Jy\kms, 
and it is more realistic to place the clouds at a distribution of distances rather than an average distance.   For instance one
of the larger clouds which is part of Complex A has
a flux of 52,210 Jy \kms~and is known to be between $4.0-9.9$ kpc (van Woerden et al. 1999).
%, giving this cloud of Complex A a mass range of $2 - 12 \times 10^{5}$ \Msun.
In addition, total individual cloud masses between $10^{5-7}$ \Msun\ are considered the most likely 
initial mass for each cloud in the models given the constraints on the formation of the clouds (i.e., the ability
to fragment and the conduction limit), cloud survival (i.e., Kelvin-Helmholtz Instability), and
cloud motion (i.e., ambient drag) (MB05; Kaufmann et al. 2005).  
Placing the clouds at a random distribution of distances $<150$ kpc and ensuring that the total mass of each
individual cloud remains below $10^{7}$ \Msun\ (again with the HI 10.5\%
of the total mass) results in a range
of total masses in condensed clouds between $1.1 - 1.3 \times 10^{9}$ \Msun.    
All of the range of masses are presented at the 95\% confidence level and 
were found after running over 100 random distributions of distances.
The lower limit on the mass of inidvidual clouds is left open, as the clouds are expected
to be disrupted as they come into close proximity with the Galaxy.
An example of the distribution
of distances and total individual cloud masses generated is shown in Figure 3.  If we adopt the
tighter distance constraint of $< 60$ kpc for all clouds, this range of total mass decreases to $4.5 - 5.7 \times 10^{8}$ \Msun.
If all of the clouds were actually given the range of masses between $10^{5-7}$ \Msun\ and
kept within 150 kpc, the total mass in potentially condensed clouds goes up somewhat to $1.2-1.4 \times 10^{9}$ \Msun.
If all clouds are given the same mass there is no direct correlation between the
clouds' resulting distances and their observed GSR velocities.  This might be expected
if the small clouds represent distant clouds not yet affected by the Galaxy's halo medium
or gravitational pull.

Clouds that have not yet been detected by existing surveys will have low HI fluxes and most
likely small masses, but if numerous they could significantly change the total mass in condensed clouds in
the halo. 
The effect these yet undetected clouds may have on the total mass has been tested by extrapolating the HVC flux 
distribution function of log N(S) $= -1.44$ log S + 3.91 (Wakker 2004), where N(S) is the number of clouds with a given
flux S, measured in 10 Jy \kms~bins.  This distribution holds to approximately 25 Jy \kms~for the updated 
WvW91 catalog and the fact that it does not continue further is at least partially due to the completeness 
limits of the catalog,
as also found in the southern HIPASS HVC catalog (Putman et al. 2002).  When the HVC flux 
distribution function is extrapolated
to continue to 1 Jy \kms~and these low flux clouds are assigned with the same random range of distances within 150 kpc, the
total mass in clouds increases only slightly to $1.1 - 1.4 \times 10^{9}$ \Msun.  
Keeping the clouds within 60 kpc and adopting this flux distribution function results in a total mass 
range of $4.5 - 6.1 \times 10^{8}$ \Msun.
Thus, with a fixed ionized component, the potentially existing lower flux clouds 
will not add significantly to the total mass in condensed clouds.
% unless they represent the intial condensed seeds
%within a primarily ionized cloud.  
If a model is adopted in which the smallest flux
clouds ($< 100$ Jy \kms; a flux cut that encompasses the majority of the compact HVCs; Putman et al. 2002) 
are at closer distances ($< 20$ kpc) and the rest of the HVC population extends to 150 kpc, the
range of total mass in condensed clouds is $0.9 - 1.1 \times 10^{9}$ \Msun. 
If on the other hand the clouds with small HI fluxes ($< 100$ Jy \kms~again) are presumed to be at
distances greater than 60 kpc (but $< 150$ kpc) the total mass in condensed clouds increases to $1.3 - 1.6 \times 10^{9}$ \Msun.
Finally, if all of these clouds were actually given a mass range between $10^{5-7}$ \Msun\ (somewhat unrealistic, as it places the
clouds with fluxes of 1-2 Jy \kms\ between 150-200 kpc), the total mass reaches a similar 
$1.4-1.6 \times 10^{9}$ \Msun.  Keeping
all of the clouds within 150 kpc requires lowering the bottom of the cloud mass range to $5 \times 10^{4}$ \Msun\ and the total mass
in condensed clouds lowers to $1.3-1.5 \times 10^{9}$ \Msun.  
%In any case, total mass is less than 2x 109 in any distribution.
%With all of the total mass ranges it is useful to remember that
%the masses represent 10.5\% neutral hydrogen (70\% hydrogen and 15\% of that hydrogen neutral).

\section{Discussion}

The total mass of potentially condensed clouds in the Galactic halo is limited to be at most approximately 1.6 $\times$
$10^{9}$ \Msun\ in all of the above cloud distributions that are consistent with the distance constraints and 
keeping 15\% of the hydrogen in each cloud neutral.   This upper mass limit is set by extrapolating the flux distribution function to 
encompass yet undetected HI clouds, placing the clouds at a random distribution
of distances below 150 kpc, and constraining their individual total mass to be below $10^{7}$ \Msun.
This mass is over a factor of 2 lower if the clouds are kept within 60 kpc.
The 3 main factors affecting this total mass are the distances, the limit on the total mass of
individual clouds, and the percentage of the cloud that is neutral.  Each of these factors
are discussed here, followed by a discussion of the impact of these results on the gas accretion
models.

\subsection{Distances to HVCs}
As discussed in the introduction, there are now a large number of constraints on the
distances to HVCs.  All of the constraints are consistent with placing the clouds within
150 kpc and some of the constraints place the clouds at closer distances.  
%For example, direct 
%distance constraints place a few clouds within 20 kpc (Wakker 2001;
%Thom et al. 2006)\footnote{One should note that there are several $>$ a few kpc direct 
%distance constraints as well.}, and if M31 can be inferred to have a similar population of HVCs as the Galaxy,
%the recent work of Westmeier et al. (2005) indicates that even the
%compact HVCs are not found beyond 60 kpc.
If the HVCs extend only out to 60 kpc and not 150 kpc and the clouds
are kept at 15\% ionized, the total mass in condensed clouds changes from $1.1-1.4 \times 10^{9}$ \Msun\ 
to $4.5 - 6.1 \times 10^{8}$ \Msun.  If we refer back to Figure 2, this type of total mass would place the majority
of the HI flux in the halo
at an average distance of approximately 21 kpc or 12 kpc, respectively.  
In either case, the upper limit on the total mass in condensed clouds is dependent
on how many clouds are at largest distances.  This will be constrained further with
ongoing distance determination programs (e.g., Thom et al. 2006).   The limit on the total mass in
clouds if they are within 60 kpc is consistent with the findings for M31 
of  $3-4 \times 10^{8}$ \Msun~(assuming 10.5\% HI again) in clouds
within 25 kpc of this galaxy (Thilker et al. 2004).
Closer distances for the clouds in the models may infer the cooling times are longer than initially presumed
and/or the densities of the fragments in the outer Galactic halo are not high enough.

\subsection{Individual Cloud Masses}
The mass of individual clouds as the condense within the hot halo depends on a number
of factors and is not currently tightly constrained.  MB04 find a range of masses are suitable, 
with $10^{5-7}$ \Msun\ being the most likely
given the constraints of conduction, evaporation, Kelvin-Helmhotz instability, and ram pressure drag.
Kaufmann et al. (2005) and Sommer-Larsen (2006) find a similar range of masses for
individual clouds.
  Tidal disruption is a factor that could disrupt the clouds within approximately
 13 kpc and lead to some small clouds that no longer have typical masses in this range (MB04).
$10^{7}$ \Msun\ is therefore adopted as the upper limit on the total mass of the individual
clouds, and since some HVCs may lie at distances below 13 kpc and represent condensed
clouds that have been disrupted, the lower limit on the mass of individual clouds is
left open, as indicated in Fig. 3.  If the 
small clouds, or lower flux clouds, are presumed to lie at distances below 20 kpc (rather
than extending out to 150 kpc) while the rest of the 
clouds are allowed to extend out to 150 kpc (as
long as their mass remains below $10^{7}$ \Msun) the total mass in clouds drops to $0.9 - 1.1 \times 10^{9}$ \Msun.
One could also argue that the small clouds should be more distant however; simply based on their angular size.
Placing the small clouds at distances greater than 60 kpc, but within 150 kpc,
increases the total mass in condensed clouds to $1.3-1.6 \times 10^{9}$.    The middle ground of placing the
clouds at a range of distances appears to be the best approach.
In contrast to the small clouds, the largest HVC complexes are unlikely to be
beyond 20 kpc, and this is kept the case by constraining a cloud's total mass to be below $10^{7}$ \Msun.

%If all of the clouds are given the same mass results in no correlation between their distance 
 %and GSR velocity.  
 %This argues against the clouds all having
 %similar masses, as the clouds at larger distances are likely to have larger infall velocities than 
 %those that are nearby given the drag force of the remaining hot halo.  ACTUALLY THIS COULD GO EITHER WAY, THEY WOULD
 % BE SLOWED AS THEY ARE ASSIMILATED INTO THE DISK AND BEGIN ROTATING WITH THE HALO
 %The lower GSR clouds could be the lower mass clouds which are more susceptible to this drag.
%K05 predict the velocity distribution of infalling, condensed gas as a function of
%height above the Galactic disk and find the gas above the disk rotates more slowly, as observed
%for some galaxies (e.g., Fraternali et al. 2002).  They also predict that it would be infalling with a velocity 
%of only 10~\kms~due to the drag exerted by the diffuse corona.  As shown in Fig.~\ref{hidist} many of the HVCs 
%are likely to have much higher infall
%velocities than this.  This discrepancy may be due to the K05
%model only resolving clouds at distances out to 30 kpc from the disk.

\subsection{Neutral Fraction}
This paper uses the nominal value of 15\% of the hydrogen in each cloud being neutral based
on the impact of the ionizing flux from the extragalactic background light ($10^{4}$ photons cm$^{-2}$ s$^{-1}$;
Maloney \& Putman 2003) and the majority of the clouds lying at distances
between 60-100 kpc.  
%MB04 suggest this would be the dominant component in keeping the clouds ionized.  
The possibility that a larger percentage of most clouds are ionized should be considered.
%The alternative of more of the cloud being neutral could also be considered, but since this would
%simply bring the total mass in potentially condensed clouds even lower than that expected 
%in the models, we concentrate on the former.
A larger ionized component could be due to many of the clouds lying at closer distances and being
subject to the ionizing radiation
from our Galaxy (e.g., Putman et al. 2003a) or via collisional ionization as
the HVCs interact with the remaining hot halo medium (e.g., Sembach et al. 2003).   It is
also possible some of the small clouds are more distant and a lower fraction of the cloud has
cooled to be observable in HI.   If only 1\% of the hydrogen in each 
cloud is detectable in HI, the total mass of clouds in the halo would reach approximately
$1.7 - 2.1 \times 10^{10}$ \Msun\ with the clouds within 150 kpc and $6.8 - 9.2 \times 10^{9}$ \Msun\ with the clouds
within 60 kpc.  A 1\% neutral component is possible for some
clouds given the discovery of highly ionized HVCs (Sembach
et al. 1999; Fox et al. 2005), but is unlikely for the majority of the HVCs given the results of 
pointed H$\alpha$ measurements (e.g., Tufte, Reynolds \& Haffner 1998; Putman et al. 2003a).  In fact 
the current limited H$\alpha$ measurements generally show a larger fraction of neutral material
than ionized for HI HVCs.
There are some HI HVCs that show evidence for extended ionized components (e.g., Sembach et al. 2003; Haffner 2005), 
but others do not (Putman et al. 2003a).   Though proximity to the Galaxy
may result in more of a cloud being ionized, this may also be offset by the
clouds closer to the Galaxy harboring higher density material. 
Estimating the actual fraction of ionized material relative to neutral is difficult given the 
limited pencil beam sightlines used to probe the ionized component. 
15\% neutral is a reasonable estimate based on the theoretical predictions and current observational constraints.
Future sensitive Fabry-Perot H$\alpha$ observations should help to clarify the full extent of the 
ionized component of HVCs.
%seems like you should be able to do more with this given that it has a large effect on the mass
%OVI results, at 70 kpc, mass is 4 x 10^{8}, at 70 kpc, the HI (10^{10} x .105) is 10^{9}, most OVI is associated
%with known complexes, but this is Ken assuming it is evenly distributed in the halo
%

The fraction of each cloud that is neutral will also depend
on the amount of metals in the gas.  As with the direct distances, there are very few HVC
metallicity determinations.  The metallicity determinations for the clouds in the sample
presented here are almost solely limited to the giant Complex C and generally range from 0.1 - 0.3 solar (e.g., Collins et al. 2002; Tripp et al. 2003).
%K05 began to explore the effect of including metals into the cooling halo gas and found
%that a metallicity of this  type doubled the energy loss due to cooling and increased the total mass in clouds.
%This increase in the total mass of clouds does not help the fact that the models already over-predict
%the mass in clouds in the halo.   
The model of MB04 uses a metallicity of 0.1 solar, so consistent with, but
on the low end of the limited HVC metallicity estimates.   If they included
more metal-rich gas, the gas would
cool more efficiently at lower densities and larger radii from the Galaxy.  
The HVCs would then be expected to lie at even larger distances ($\sim 200$ kpc)
which seems unlikely given the HVC distance constraints.  Higher metallicities in the MB04 
model would also most likely result in an increased total mass of condensed halo clouds.
%, and an even large discrepancy between the observations and models.

\subsection{Implications on Gas Accretion} 
Disk galaxy formation models generally predict the existence of a hot halo with
a baryonic mass of a few $\times\ 10^{10}$ \Msun~ (e.g., MB04; Fukugita \& Peebles 2005; Sommer-Larsen 2006).
 Hot halo gas has recently been detected in the vicinity of our Galaxy (Sembach et al. 2003; Rasmussen et al. 2002) and
around other spirals (Pedersen et al. 2006) in possible support of these models.
%; however the current mass estimates for the hot gas around the Galaxy are
%an order of magnitude below the model predictions.
This hot gas gradually cools as clouds and fuels the galaxy's star formation, but the specifics
of the process vary by model.  The model of MB04 predicts that
cloud formation and infall balances at early times, with the
balance of condensed clouds in the halo approximately 2 $\times\ 10^{10}$ \Msun.
The analysis of the observed HI halo clouds presented here indicates the total mass in potentially 
condensed clouds in the halo is at least an order of magnitude below this.   The $< 6 \times 10^{8}$ \Msun~
range found here for the halo clouds within 60 kpc is consistent with the simulations
of Kaufmann et al. (2005) and Sommer-Larsen (2006).  
The reason for less mass in condensed halo clouds than found by MB04 may be due to 
the clouds falling in rapidly after they are formed, and thus less clouds are 
visible in the halo at a given time. 
If clouds fall in rapidly, feedback from the Galaxy may need to be considered to keep most of the halo material 
 in a warm-hot phase and not over-produce
 the baryonic mass of the Galaxy.
A Galactic fountain is one possible feedback mechanism in which the hot gas from multiple supernovae
rises into the halo (e.g., Bregman 1980).   The Galactic fountain 
%has difficulty explaining the velocity,
%spatial, and metallicities of the HVCs (refs), but 
could inject a large amount of enriched hot
gas into the lower Galactic halo which may then mix with the massive hot halo
material and cool as the intermediate velocity clouds (IVCs) found around our Galaxy.  
IVCs are much closer to disk (0.5 - 2 kpc) than
HVCs and are also of higher metallicity (0.5-2 solar; Wakker 2001).  

%Another factor that may lead to a lower total mass in condensed clouds may be a lower cooling efficiency
%in the halo than assumed in the models.  
%This is unlikely to be due to the metallicity the models assume, which could potentially
%be too low (see above),  but could
% be related to lower average density differentiations in the halo.   
 The limited mass in condensed clouds forming and falling on to the Galaxy indicates 
 additional accretion methods
 are necessary to explain the Galaxy's observed baryonic mass.  
Several of these additional accretion methods are directly evident as our Galaxy destroys smaller galactic 
systems such as the Sagittarius dwarf and Magellanic
Clouds.  The Magellanic System itself will eventually bring on the order of 10$^{9}$ \Msun\ 
of HI into the Milky Way (Putman et al. 2003b).   The
Magellanic System is an example of a less frequent, large accretion event and could also potentially
disrupt the process of cloud formation at the current epoch.
Finally, though cold accretion is unlikely to dominate at low redshifts and for galaxies as massive as the
Milky Way, this method of gas accretion in
which the incoming gas is never heated to the virial temperature of the galaxy halo, may also need to be considered (e.g., Keres et al. 2005). 
In any case, in the model of Keres et al. the smaller galaxies currently being accreted by the Galaxy
obtained the bulk of their mass via cold accretion, indicating the Galaxy is indirectly obtaining mass in this
fashion.
%with a total mass (baryonic and dark) on the order of $\sim7 \times 10^{9}$~\Msun~ disrupting 
%the equilibrium or the supply of gas into the Galactic halo from the intergalactic medium beginning to be depleted. 
 
%  Smaller total amounts of condensed clouds in the Galactic halo may be the observed
% reality, but if it holds it does have an effect on the formation and evolution of the Milky Way.
 %Smaller mass clouds fall in more rapidly and the result is a larger total baryonic mass for the Galaxy.  
% For a characteristic cloud mass of $0.5 - 1 \times 10^{6}$ \Msun, the resultant baryonic mass of the Milky Way is
%approximately $7 \times 10^{10}$ \Msun, compared to the expected $4-6 \times 10^{10}$ \Msun~(Dehnen \& Binney 1998).
% MB04 note that the larger baryonic mass from the infall of smaller mass clouds is not entirely detrimental, due to 
% a hot gas core possibly contributing on the order of $4 \times 10^{10}$ \Msun.   Part of this hot core may
% be evident in recent high-velocity O VI detections tracing gas at T$\sim 10^{6}$ K (Sembach et al. 2003).   
%This then leaves the main disagreement between the model and observations as the present
%total mass of clouds in the halo, rather than the individual masses of the clouds or the resulting mass of the galaxy. 

 There are several properties of the observed HI clouds in the halo that can be used to constrain the models
 as they are developed further besides the mass limits.
As discussed in the results section and shown in Fig. 1, most of the observed clouds are found at latitudes below 60\deg\ and 
there are several clusters of smaller clouds in specific directions that may represent
the preferred accretion axes of the Galaxy.  The mean negative V$_{\rm GSR}$ of the cloud population (-43 \kms)
is suggestive of a net infall, but the mix of positive and negative V$_{\rm GSR}$ clouds
found throughout the sky suggests it is not a simple model.  Since we are only measuring one
component of the cloud's velocity, some clouds with negative or positive V$_{\rm GSR}$'s
may be moving away or towards the disk, respectively.   In any case, after forming the clouds
appear to be on a variety of orbits, which after collisions and ram pressure drag eventually
lead to infall (MB04).  There is no correlation between the GSR velocity of the cloud and 
distance if all of the clouds are given similar masses, indicating that the clouds do not
all have similar masses or reflecting the complex motions of the halo clouds.
%towards the anti-center, $l=270$\deg, and $l=45$\deg.   
 %THE NAMES OF THESE COMPLEXES AREN'T REALLY RELEVANT TO THE MODELS
 %These clusters contain predominantly small clouds that are
 %considered part of the the Anti-Center Complex or one of the W complexes (CHECK MAP, MAYBE
 %GALACTIC CENTER TOO).  
 If the ongoing distance determination programs continue to place the HVCs within 60 kpc
 of the Galactic disk, explanations will need to be made for the reason the HI is only found
 out to this radii and what percentage of the baryonic mass the HI represents.  
% Observationally the clouds potentially have a wide range of masses depending on each
% clouds distance, but the constraints thus far are consistent with the clouds having masses less than $10^{6}$ \Msun\ in
% HI.  
Future HI surveys being completed by the Galactic Arecibo L-Band Feed Array (GALFA) consortium (e.g., Stanimirovic et al. 2006) and
the Galactic All-Sky Survey (GASS; McClure-Griffiths et al. 2006), will be important for characterizing the properties of the clouds
and their relationship to the Galactic disk.  
The GALFA surveys, with their increased sensitivity and resolution, will be particularly important
for examining the flux distribution function and assessing
%It is likely that more compact clouds or lower column density clouds will be detected by these
%surveys which will slightly increase the total mass in halo clouds and possibly represent some of the mostly
%ionized clouds just beginning to condense.  The Arecibo surveys will also allow for an analysis of 
how the condensed clouds interact with the diffuse hot halo as they are assimilated into
the Galactic disk.  This analysis will also allow for an estimate of the mass of the elusive diffuse hot halo.

%  Is there a way to simply increase the total observed mass of potentially condensed clouds in the halo?
% One can either increase the mass of the individual clouds or increase the total number of clouds.
% Since based on distance constraints, the mass cannot be increased by placing the clouds at
% greater distances, increasing the total
%masses of the individual clouds would require additional ionized material or lower column density material
%which has not yet been observed.  This may be achieved if the timescales for the cooling of the gas are comparable
%to the infall times and the contribution from the extragalactic ionizing radiation field is substantial.  
% Lower column density extensions of clouds have been detected by pointed deep observations into the Galactic
% halo (to $\sim 8\times 10^{17}$ cm$^{2}$; Lockman et al. 2002) and more extensions may
 %be revealed with future surveys.   
%Increasing the total number of clouds may be achieved by many of the same factors that increase the
%mass of the individual clouds.  Some clouds may not have
%had time to cool to be observable as HVCs and be fully ionized.  
%It is also possible that some intermediate velocity clouds (IVCs) respresent the condensed clouds at
%their final stage of accretion.  Given the differences in metallicity between some IVCs and HVCs, as mentioned
%above, it is unlikely the majority of the IVCs fit naturally into this model, but if they
%were all considered they could potentially increase the mass estimates by a factor of 2 (Wakker \& van Woerden 1997).
 
\acknowledgments
Thanks to James Bullock, Ari Maller, and Jesper Sommer-Larsen for very useful discussions, to the referee for insightful comments, and to Bart Wakker for providing an updated version of the WvW91 catalog.

\clearpage

\clearpage

\clearpage

\begin{figure}
\includegraphics[angle=90, scale=0.8]{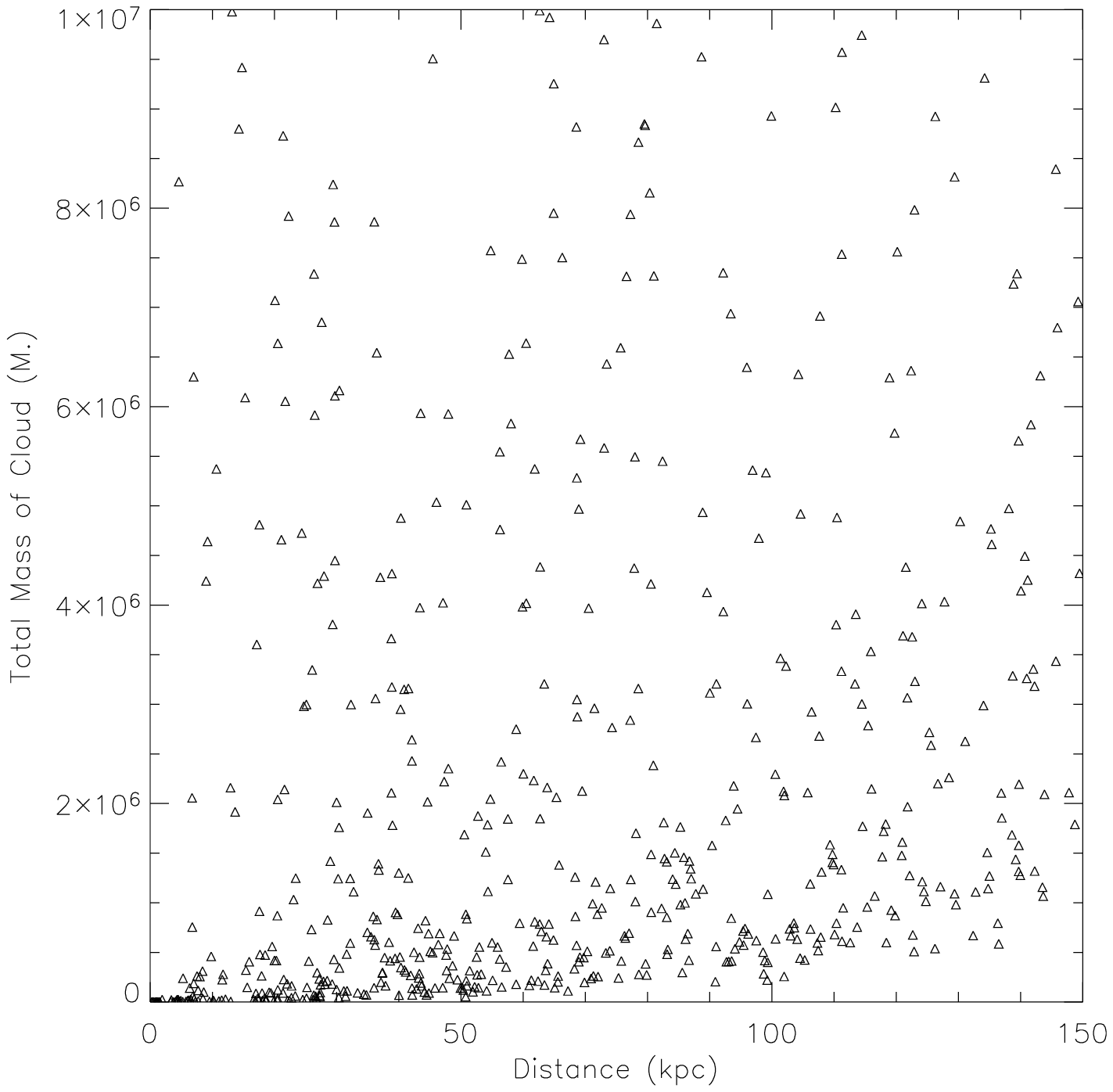}
\caption{An example of the total masses of individual clouds if they are assigned
a random distribution of distances below 150 kpc, are 10.5\% observable neutral hydrogen,
and are confined to have a total mass below $10^{7}$ \Msun.   The total mass in
condensed clouds with this data is typically $1.1-1.3 \times 10^{9}$ \Msun.}
\end{figure}


\begin{thebibliography}{}
\bibitem{} Bregman, J.N. 1980, ApJ, 236, 577
\bibitem{} Br\"uns, C. \& Westmeier, T. 2004, A\&A, 426, 9
%\bibitem{} Br\"uns, C., Kerp, J., Staveley-Smith, L., Mebold, U., Putman, M.E., Haynes, R.F., Kalberla, P.M.W., 
%Muller, E., Filipovic, M.D. 2005, A\&A, 432, 45
%\bibitem{} de Blok, W.J.G., Zwaan, M.A., Dijkstra, M., Briggs, F.H. \& Freeman, K.C. 2002, A\&A, 382, 43
\bibitem{} Collins, J., Shull, M. \& Giroux, M. 2003, ApJ, 585, 336
%\bibitem{} Dehnen, W. \& Binney, J. 1998, MNRAS, 294, 429
\bibitem{} Fox, A.J., Wakker, B.P., Savage, B.D., Tripp, T.M., Sembach, K.R., Bland-Hawthorn, J.  2005, ApJ, 630, 332
\bibitem{} Fraternali, F., van Moorsel, G., Sancisi, R. \& Oosterloo, T.  2002, AJ, 123, 3124
\bibitem{} Fukugita, M. \& Peebles, P.J.E. 2005 (astro-ph/0508040)
\bibitem{} Haffner, L.M. 2005, in Extra-Planar Gas, ASP Conf. Proc. V. 331, ed. R. Braun, 25
\bibitem{} Hulsbosch, A.N.M. \& Wakker, B.P. 1988, A\&AS, 75, 191
\bibitem{} Kauffman, T., Mayer, L., Wadsley, J., Stadel, J. \& Moore, B. 2005, MNRAS, submitted (astro-ph/0507296)
\bibitem{} Keres, D., Katz, N., Weinberg, D.H., \& Dav\'e, R. 2005, MNRAS, 363, 2
%\bibitem{} Lockman, F.J., Murphy, E.M., Petty-Powell, S., \& Urick, V.J. 2002, ApJS, 140, 331
\bibitem{} McClure-Griffiths, N. et al. 2006, ApJ, 638, 196
\bibitem{} Maller, A.H. \& Bullock, J.S. 2004, MNRAS, 355, 694
\bibitem{} Maloney, P.R. \& Putman, M.E. 2003, ApJ, 589, 270
\bibitem{} Morras, R., Bajaja, E., Arnal, E.M., Poppel, W.G.L. 2000, A\&AS, 142, 25
\bibitem{} Oort, J.H., 1966, Bull. Astron. Inst. Netherlands, 18, 421
\bibitem{} Pedersen, K., Rasmussen, J., Sommer-Larsen, J. Toft, S., Benson, A.J., Bower, R.C. 2006, New Astronomy, in press (astro-ph/0511682)
\bibitem{} Pisano, D.J., Barnes, D.G., Gibson, B.K., Staveley-Smith, L., Freeman, K.C.,  \& Kilborn, V. 2004, ApJ, 610, L17
\bibitem{} Putman, M.E. et al. 2002, AJ, 123, 873
\bibitem{} Putman, M.E., Bland-Hawthon, J., Veilleux, S., Gibson, B.K., Freeman, K.C. \& Maloney, P.R. 2003a, ApJ, 597, 948
\bibitem{} Putman, M.E., Staveley-Smith, L., Freeman, K.C., Gibson, B.K., \& Barnes, D.G. 2003b, ApJ, 586, 170
\bibitem{} Renda, A., Kawata, D., Fenner, Y. \& Gibson, B.K. 2005, MNRAS, 356, 1071
\bibitem{} Rasmussen, A., Kahn, S., Paerels, F. 2003, in The IGM/Galaxy Connection, eds. J. Rosenberg \& M. Putman, Kluwer, 281, 109
\bibitem{} Rocha-Pinto, H.J., Scalo, J., Maciel, W.J., \& Flynn, C. 2000, A\&A, 358, 869
\bibitem{} Sembach, K.R., Savage, B.D., Lu, L., \& Murphy, E.M. 1999, ApJ, 515, 108
\bibitem{} Sembach, K.R. et al. 2003, ApJS, 146, 165
\bibitem{} Sommer-Larsen, J. 2006, ApJL, submitted (astro-ph/0602595)
\bibitem{} Stanimirovic, S., Putman, M.E., Heiles, C., Goldston, J. et al. 2006, ApJ, in preparation
\bibitem{} Thilker, D. et al. 2004, ApJL, 601, 39
\bibitem{} Thom, C., Putman, M.E., Gibson, B.K., Christleib, N., Flynn, C., Beers, T.C., Wilhelm, R., Lee, Y.S. 2006, ApJL, 638, 97
\bibitem{} Tripp, T. et al. 2003, AJ, 125, 3122 
\bibitem{}Tufte, S.L., Haffner, L.M., \& Reynolds, R.J. 1998, ApJ, 508, 200
\bibitem{} van Woerden, H. et al. 1999, Nature, 400, 138
%\bibitem{} Tufte, S.L., Wilson, J.D., Madsen, G.J., Haffner, L.M. \& Reynolds, R.J. 2002, ApJ, 572, 153
\bibitem{} Wakker, B.P. 2004, in High-Velocity Clouds, eds. H. van Woerden, B.P. Wakker, Schwarz, U.J., \&
de Boer, K.S., Kluwer, 312, 25
%\bibitem{} Wakker, B.P. et al. 2003, ApJS, 146, 1
\bibitem{} Wakker, B.P. 2001, ApJS, 136, 463
\bibitem{} Wakker, B.P. \& van Woerden, H. 1997, ARA\&A, 35, 217
\bibitem{} Wakker, B.P. \& van Woerden, H. 1991, A\&A, 250, 509 (WvW91)
\bibitem{} Westmeier, T., Br\"uns, C. \& Kerp, J.  in Extra-Planar Gas, ASP Conf. Proc. V. 331, eds. R. Braun, 105
\bibitem{} White, S.D.M. \& Rees, M.J. 1978, MNRAS, 183, 341
\bibitem{} White, S.D.M. \& Frenk, C.S. 1991, ApJ, 379, 52
\bibitem{} Zwaan, M. 2001, MNRAS, 325, 1142
\end{thebibliography}
\end{document}